\documentclass[aps,pra,reprint,showpacs,groupedaddress,longbibliography]{revtex4-1}

\usepackage{empheq} 
\usepackage{mathrsfs} 
\usepackage{mathtools,amssymb,graphicx,units}
\usepackage{multirow}
\usepackage[plainpages=false,pdfpagelabels,colorlinks=true,linkcolor=red,urlcolor=blue,citecolor=blue,pdftitle={Title},pdfauthor={},pdfdisplaydoctitle=true,pdfduplex=DuplexFlipLongEdge]{hyperref}
\usepackage{soul} 
\usepackage[dvipsnames,usenames]{color}
\usepackage[normalem]{ulem}
\usepackage{graphicx}
\graphicspath{{figures/}}
\usepackage{xcolor}
\usepackage{bbold} 
\usepackage{float}
\emergencystretch=\maxdimen
\def\ave#1{\langle #1\rangle}
\newcommand{\bra}[1]{\langle #1|}
\newcommand{\ket}[1]{|#1\rangle}
\newcommand{\braket}[2]{\langle #1|#2\rangle}

\newcommand{\up}{\!\uparrow}
\newcommand{\dn}{\!\downarrow}

\newcommand{\Sv}{\mathbf{S}}

\newcommand{\calH}{{\mathcal H}}

\newcommand{\scrS}{\mathscr{S}}
\newcommand{\scrC}{\mathscr{C}}

\newcommand{\hf}{\frac{1}{2}}

\begin{document}

\title{
Four interacting spins: addition of angular momenta,\\ spin-spin correlation functions, and entanglement}
\author{Raimundo R. dos Santos} 
\author{Lucas Alves Oliveira} 
\altaffiliation{Current address: Departamento de F\'\i sica, PUC/Rio}
\author{Natanael C. Costa} 
\affiliation{Instituto de F\'\i sica, Universidade Federal do Rio de Janeiro,
21941-972 Rio de Janeiro RJ, Brazil}

\begin{abstract}
We study four spins on a ring coupled through competing Heisenberg interactions between nearest neighbors, $J$, and next-nearest neighbors, $J_2\equiv\alpha J>0$. 
The spectrum is obtained in a simple way by using the rules for addition of 4 angular momenta. This allows us to follow the evolution of the ground state with $\alpha$, characterized by level crossings and by analyses of spin-spin correlation functions. Further insight is obtained by examining the entanglement between different parts of the system: we observe that the entanglement entropy is strongly dependent on how the system is partitioned.

\end{abstract}

\maketitle


\section{Introduction}
\label{sec:intro}

Some many-body systems are characterized by the presence of a quantum critical point (QCP) occurring at zero absolute temperature, which generically separates an ordered (or quasi-ordered) phase from a disordered one. 
This phase transition is driven by some control parameter \cite{Sachdev11,Continentino17}, which for magnetic systems may be an external transverse field, competing interactions, doping fraction, pressure, and so forth. 
Despite being a zero temperature (that is, ground state) phenomenon, the presence of a QCP influences the behavior of measurable quantities at finite temperatures \cite{Sachdev11,Continentino17}. 
Given that the singularities appearing in second order phase transitions only set in in the thermodynamic limit, a widely used theoretical strategy to study these phenomena is to extract information from small-sized systems and use finite-size scaling ideas \cite{Fisher71,Barber83} to predict the large system behavior.    
To this end, at zero temperature we traditionally have at our disposal properties such as spectral gaps, response functions, and correlation functions; further, in recent years entanglement measures, which have been at the heart of proposals for quantum computation \cite{Nielsen10,Chitambar19}, have been used as signatures of quantum critical behavior~\cite{Osterloh02}.

While one usually resorts to numerical techniques to calculate these properties for systems ranging from typically tens to hundreds of spins, the possibility of obtaining them analytically for just a few spins proves extremely useful. 
Indeed, this leads to crucial insights from exploring symmetries which render calculations with great simplicity, while it may provide data to check the numerical codes devised for larger system sizes.
It also sheds light into exploring less familiar probes, such as entanglement.
From the pedagogical point of view, dealing with few spins provides examples of how to add more than two angular momenta in a systematic way. 

\begin{figure}[t]
	\vskip 0.5cm
	\centering\includegraphics[scale=0.35]{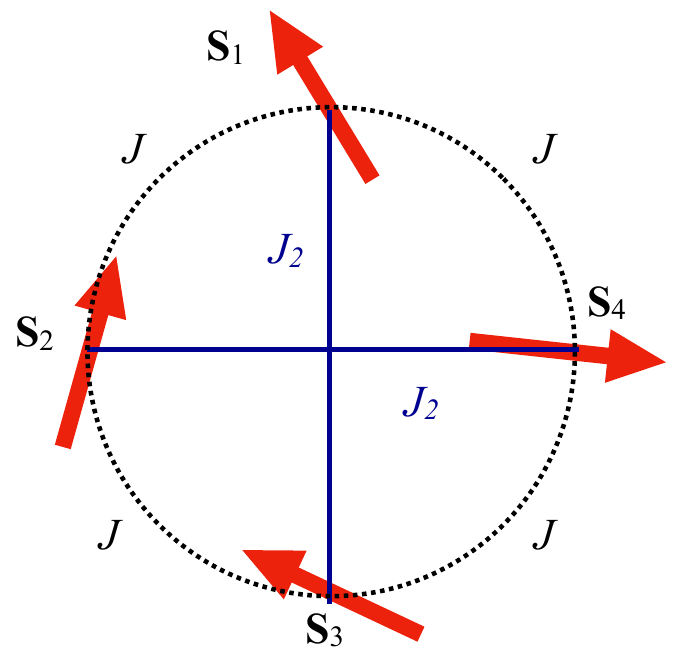} 
	\caption{A semiclassical representation of the 4 spins on a ring; $J$ and $J_2$ are the exchange coupling constants between spins on nearest (dotted lines) and next-nearest neighbor sites (full lines), respectively.}
\label{fig:4spins}
\end{figure}

Within this context, a particularly interesting example of such a system consists of four spins-1/2 fixed in positions on a ring; see Fig.\,\ref{fig:4spins}. 
We assume they are coupled through competing exchange interactions $J$ and $J_2=\alpha J$ between nearest- and next-nearest neighbors, respectively; quantum effects are brought about by considering scalar interactions involving the three cartesian spin components.    
The Hamiltonian may therefore be expressed as
\begin{equation}
	\mathcal{H} = J \left[\sum_{i=1}^4 \Sv_i\cdot\Sv_{i+1}+\alpha\sum_{i=1}^4 \Sv_i\cdot\Sv_{i+2}\right],
\label{eq:HJ1J2}	
\end{equation}
with $\Sv_{5}\!\equiv\!\Sv_1$ and $\Sv_{6}\!\equiv\!\Sv_2$, thus setting up periodic boundary conditions, as in Fig.\,\ref{fig:4spins};
for reasons which will become apparent soon, we always consider $J_2\geq0$, so that the sign of $\alpha$ is the same as that of $J$.
A semi-classical analysis for $\alpha=0$ immediately reveals that if $J<0$ the ground state corresponds to all spins aligned (ferromagnetic state),  e.g.\ $\ket{\!\up \up \up \up}$, while if $J>0$ the tendency is to have two interpenetrating sublattices, each with aligned spins, but aligned antiparallel with each other [antiferromagnetic (AFM), or N\'eel-like, state], e.g.\ $\ket{\!\!\up \dn \up \dn}$.
Therefore, when $\alpha>0$ a large $J_2$ disrupts the AFM state, favoring a dimerized state, $\ket{\!\!\up \dn \dn \up}$; by the same token, when $\alpha<0$,  a large $J_2$ also favors a dimerized state. 
The reader should note that when $J_2<0$ there is no competition with $J$, irrespective of the sign of the latter.
An interesting issue is to determine the ground state evolution as $J_2$ increases from zero, for both $\alpha>0$ and  $\alpha<0$.  
Here we will use addition of four angular momenta to obtain simple expressions for both the eigenstates and energies for $\mathcal{H}$, from which we discuss the different `phase transitions' based on level crossings, correlation functions, and entanglement entropy.
As discussed later, this example with four spins allows a much more diverse and illustrative analysis of entanglement than with just two spins.

The layout of the paper is as follows. 
In Sec.\,\ref{sec:spectrum4} we present a diagonalization of $\mathcal{H}$ based on the addition of four spins-1/2, and discuss its spectrum.  
This solution allows us to obtain in Sec.\,\ref{sec:corrfns} the spin-spin correlation functions for the different ground states, when we also highlight their physical content. 
In Sec.\,\ref{sec:Entangle4} we obtain the entanglement entropies for the different ways we can split the system in two parts; the dependence with $\alpha$ also reveals their ability of pinpointing QCP's.  
In Section \ref{sec:larger} we briefly discuss the main features of this model, as unveiled by calculations on larger systems.
And, finally, Sec.\,\ref{sec:concl} presents our  conclusions.
 \color{black}


\section{The spectrum}
\label{sec:spectrum4}
We start by recalling the notation relative to eigenstates of the spin operator for particle $i$, $i=1,...4$, 
\begin{align}
	\Sv_i^2\ket{s_im_i} &= s_i(s_i+1)\hbar^2\ket{s_im_i},\\
	S_i^z\ket{s_im_i} &= m_i\hbar\ket{s_im_i}. 
\label{eq:Seketsi}
\end{align}	
Since we only consider here spin-1/2 particles, $s_i=~s=~1/2, \forall i$, and $m_i=\pm1/2$ (or $\up$, $\dn$, for short), $\forall i$. 
A possible basis for this 16-dimensional state space is provided by $\{\ket{s_1m_1s_2m_2s_3m_3s_4m_4}\}$. 
Further, since the labels $s_i$ reflect an intrinsic attribute of the particles, they remain fixed and may be omitted to simplify the notation; that is, we refer to this basis as $\{\ket{m_1m_2m_3m_4}\}$.
In this basis, the Hamiltonian appears in block diagonal form, each characterized by the value of $M=\sum_{i=1}^4 m_i$; this block structure reflects the built-in axial symmetry of this basis due to the conservation of the $z$-component of the total spin, $S_T^z\equiv S_1^z+S_2^z+S_3^z+S_4^z$.

However, the Hamiltonian symmetry is actually higher than this: it is invariant under a simultaneous rotation of \emph{all} spins by any angle around \emph{any} axis.
That is, the total angular momentum is conserved, as it can be  
recalled that classically the absence of an external torque leads to the conservation of the total angular momentum vector.
Nonetheless, quantum-mechanically the non-commutation of the components of angular momentum operators does not allow their simultaneous determination.           
At any rate, further simplicity should arise by changing to a higher-symmetry basis, e.g.\ to one labelled by the total angular momentum quantum number, $S_T$, in addition to $M$, which is the quantum number associated with $S_T^z$. 

The total spin operator is
\begin{equation}
	\Sv_T = \Sv_1+ \Sv_2+ \Sv_3+ \Sv_4,
\end{equation}	
whose square may be written as 
\begin{equation}
	\Sv_T^2 = K_0 + 2 K_1 +  K_2,
\label{eq:S2K}	
\end{equation}
where we have introduced 
\begin{align}
	\label{eq:K0}	
	K_0  & \equiv \sum_{i=1}^4\Sv_i^2,\\
	\label{eq:K1}
	K_1  & \equiv \sum_{i=1}^4 \Sv_i\cdot\Sv_{i+1},	\\
	\label{eq:K2}
	K_2  &\equiv \sum_{i=1}^4 \Sv_i\cdot\Sv_{i+2}.
\end{align}
Thus, the Hamiltonian, Eq.\,\eqref{eq:HJ1J2}, may be expressed as
\begin{equation}
	\mathcal{H} = J[K_1 + \alpha K_2],
\label{eq:HK1K2}	
\end{equation}	
where we note that periodic boundary conditions for 4 sites imply that the coupling between any two second-neighbor spins appears twice in the Hamiltonian.

\begin{table}
\centering
\begin{tabular}{|c|c|c|c|c|c|c|}   
\hline
         $s_{13}$ & 0 & 1 & 0 & 1 & 1 & 1  \\  
\hline
	$s_{24}$ &  0 & 0 & 1 & 1 & 1 & 1  \\
\hline
	$S$ &  0  & 1 & 1 & 0 & 1 & 2  \\
\hline
	$g_S$ &  1  & 3 & 3 & 1 & 3 & 5\\
\hline
\end{tabular}
\caption{Quantum numbers for the addition of four spins-1/2. 
The partial sum of two spins-1/2 yield the quantum numbers $s_{13}=0,1$ and $s_{24}=0,1$; A given pair of values of $s_{13}$ and $s_{24}$ yield the total spin quantum number as $S=|s_{13}-s_{24}|,|s_{13}-s_{24}|+1,\ldots s_{13}+s_{24}$. Since the total angular momentum is conserved, the energy degeneracy for each total spin multiplet is $g=2S+1$.}
\label{tab:spins_on_square}
\end{table}

Our aim now is to set up a basis in which $\mathcal{H}$ is expressed solely in terms of eigenvalues of the operators $K_r,\, r =0,1,2$.
According to the rules for addition of more than two angular momenta, one adds two spins at a time, and subsequently add the results; see, e.g.\,Ref.\,\onlinecite{Messiah70v2}.
We may then evaluate $K_2$ by taking the square of the partial sums $\Sv_{13}\equiv\Sv_1+\Sv_3$ and $\Sv_{24}\equiv\Sv_2+\Sv_4$, 
\begin{align}
	\Sv_{13}^2 &= \Sv_1^2 + \Sv_3^2 + 2\, \Sv_1\cdot \Sv_3\\
	\Sv_{24}^2 &= \Sv_2^2 + \Sv_4^2 + 2\, \Sv_2\cdot \Sv_4,
\end{align}	
and adding them to obtain, with the aid of Eq.\,\eqref{eq:K2},
\begin{equation}
	K_2=\Sv_{13}^2 +\Sv_{24}^2 - K_0.
\label{eq:K2b}	
\end{equation}
Taking this into Eq.\,\eqref{eq:S2K} leads to
\begin{align}
	K_1  = \frac{1}{2}[\Sv_T^2 - ( \Sv_{13}^2 +\Sv_{24}^2 )].
\label{eq:K1b}	
\end{align}
We note that $\Sv_T^2$, $\Sv_i^2$, $i=1-4$, and $\Sv_{ii+2}^2$, $i=1,2$ are scalar operators; as such, they commute with each other as well as with $S_T^z$.
Therefore $\mathscr{C}\equiv\{\Sv_T^2,S_T^z,\Sv_{13}^2,\Sv_{24}^2, \Sv_1^2,\Sv_2^2,\Sv_3^2,\Sv_4^2\}$ forms a complete set of commuting observables (CSCO, 
see, e.g., Refs.\,\cite{Messiah70v2, Cohen-Tannoudji77}).

\begin{figure}[t]
    \hskip -0.5cm
	\centering\includegraphics[width=9.0cm]{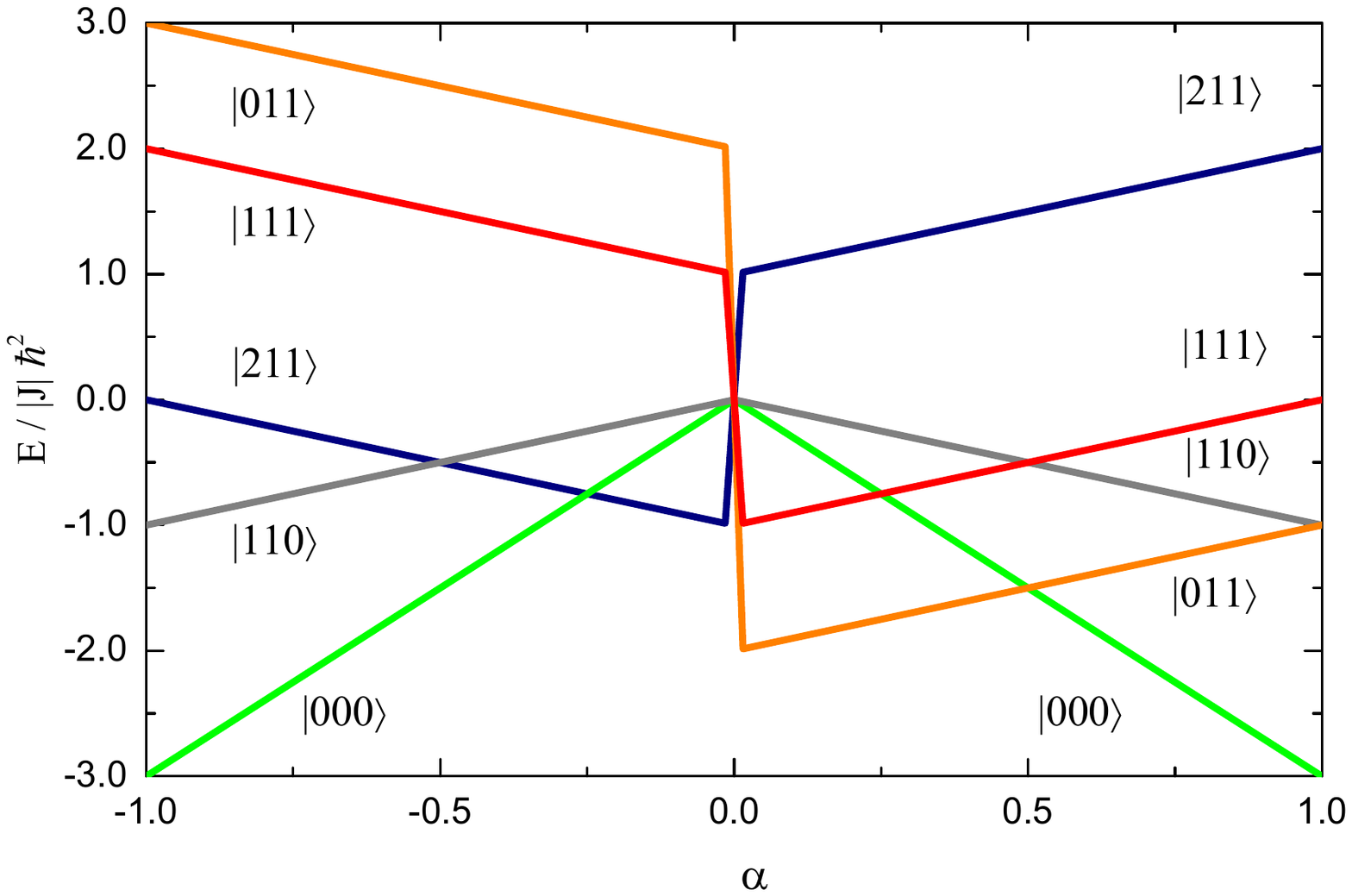}
	\vskip -0.5cm 
	\caption{Energy eigenvalues for the 4-site $J$-$J_2$ Heisenberg model as a function of $\alpha\equiv J_2/J$, with $J_2>0$. 
	The curves are labelled by a simplified notation, $\ket{S_Ts_{13}s_{24}}$, since for each $S_T$ they are degenerate in $M$; $S_T$ is the total spin quantum number, and $s_{13}$ and $s_{24}$ are the quantum numbers specifying the partial sums.
	Still within this notation, we recall that the states $\ket{110}$ and $\ket{101}$ are degenerate.
	The energies of some states display a discontinuity at $\alpha=0$, which have been purposely smoothed for easier identification. 
	}
\label{fig:Evsa4}
\end{figure}

The Hamiltonian can then be cast in the form
\begin{align}
	\mathcal{H} &= \frac{J}{2}[\Sv_T^2 - 2 \alpha K_0 + (2\alpha-1) (\Sv_{13}^2 + \Sv_{24}^2)],
\label{eq:H4final}		
\end{align}
and since \eqref{eq:H4final} is expressed solely in terms of operators in $\mathscr{C}$, we may replace the operators by their respective eigenvalues when using this basis.
The eigenvalues can be determined by systematically adding spins, starting with the partial sums, $\Sv_{13}$ and  $\Sv_{24}$: the possible quantum numbers for the square of these partial sums are $s_{ii+2} = 0,1$, $i=1,2$. 
These partial sums then add to make up the total spin, $\Sv_T =\Sv_{13}+\Sv_{24}$, whose possible quantum numbers for its square are $S_T=0,1,2$.
We finally arrive at an expression for the eigenenergies in terms of the good quantum numbers:
\begin{align}
	\frac{E}{|J|\hbar^2}&=\text{sign}(\alpha)\hf\color{black}\left\{S_T(S_T+1)- 8\alpha s(s+1)\right.\nonumber\\
	&\left.+(2\alpha-1)[s_{13}(s_{13}+1)+s_{24}(s_{24}+1)] \right\},
\label{eq:E4final}		
\end{align}
where $\text{sign}(\alpha)\equiv \alpha/|\alpha|=J/|J|$ is the sign function, which is $>0\,(<0)$ if $\alpha >0\, (\alpha<0)$.
As expected, the energies do not depend on the orientation of the ring, since there is no preferred direction in space; the degeneracy in $M=-S_T,-S_T+1,\ldots S_T$ of each level is therefore $g=2S_T+1$. 
Table \ref{tab:spins_on_square} lists all possible combinations of quantum numbers entering in the evaluation of the energies, which are plotted as functions of $\alpha$ in Figure \ref{fig:Evsa4}. 
The eigenvectors may then be denoted by $\ket{S_TMs_{13}s_{24}}$, where, for the same reasons as before, we keep omitting $s_1,\ldots,s_4$ from the list.

Figure \ref{fig:Evsa4} suggests that one should discuss the cases $\alpha<0$ and $\alpha>0$ in turn. 
In addition, it will often prove illustrative to express the lowest energy eigenstates in terms of a `split basis', formed by $\{\ket{s_{13}m_{13}}\otimes\ket{s_{24}m_{2}}\}$; see Appendix \ref{app:split}.

If $-1/4 < \alpha<0$ the ground state is the ferromagnetic quintuplet, $\ket{2M11}$, $M=-2,-1,0,1,2$; see Eqs.\,\eqref{eq:2211otimes}, \eqref{eq:2111otimes}-\eqref{eq:2-211otimes}.
By contrast, if $\alpha<-1/4$, the ground state is the singlet
\begin{equation}
	\label{eq:0000otimestxt}
		\ket{0000}=\ket{00}\otimes\ket{00}.	
\end{equation}
Thus we may say that the nature of the ground state changes at $\alpha_{c,1}=-1/4$, and it is suggestive that this evolves into a quantum critical point as the number of sites,  $N$, increases; more on this below. 
 
If $\alpha>0$, the second neighbor coupling frustrates the tendency of forming a N\'eel-like state, $\ket{\!\!\uparrow\downarrow\uparrow\downarrow}$.
Indeed, Figure \ref{fig:Evsa4} shows that if $0<\alpha < 1/2$, the ground state is a different singlet,
\begin{equation}
	\label{eq:0011otimestxt}
		\ket{0011}=\frac{1}{\sqrt{3}}[\ket{1-1}\otimes\ket{11}-\ket{10}\otimes\ket{10}+\ket{11}\otimes\ket{1-1}]
\end{equation}	
[see Eq.\,\eqref{eq:0011otimes}], which is a superposition of pairs (1,3) and (2,4) forming triplets adding in such way to yield a global singlet; note that each term in the superposition has $M=m_{13}+m_{24}=0$.

Beyond $\alpha=\alpha^*=1/2$, the ground state is the same singlet as for $\alpha<-1/4$; see Eq.\,\eqref{eq:0000otimestxt}. 
As we will see, the correlation functions highlight the difference between the two singlets appearing when $\alpha>0$.

\section{Spin-spin correlation functions}
\label{sec:corrfns}

The spin-spin correlation function is defined as 
\begin{equation}
	\label{eq:Sr}
		\scrS(r)\equiv\ave{\Sv_i\cdot\Sv_{i+r}}, 
 \end{equation}
thus measuring the influence a spin variable at site $i$ exerts on the value of the spin variable at a site separated by a distance $r$; at zero temperature, the averages $\ave{\cdots}$ are understood as ground state expectation values.

Translational invariance allows us to write
\begin{equation}
	 \label{eq:SrK}
	 	\scrS(r)=\frac{1}{4}\ave{K_r},\quad r=0,1,2,		
\end{equation}
where the $K_r$'s are given by Eqs.\,\eqref{eq:K0}-\eqref{eq:K2}, and we note that $r=2$ (we consider unit lattice spacing) is the maximum distance in this case due to periodic boundary conditions.
In a macroscopic system, the behavior at large distances, $r\gg 1$,  probes the existence of some degree of ordering, as well as its nature. 
Since here we are dealing with only 4 sites, we cannot make any claims about the long distance behavior of $\scrS(r)$.
Nonetheless, even for 4 sites $\scrS(r)$ provides useful insights into the nature of the different ground states, as we will see below.   

Similarly to the eigenenergies, the $\ave{K_r}$'s may be expressed in terms of the eigenvalues of the operators in the set $\scrC$, to yield
\begin{align}
	\label{S0}
		\scrS(0)&=s(s+1)\hbar^2,\\
	\label{S1}
		\scrS(1)&=\frac{1}{8} [s(s+1)-s_{13}(s_{13}+1)-s_{24}(s_{24}+1)]\hbar^2,\\
	\label{S2}
		\scrS(2)&=\frac{1}{4} [s_{13}(s_{13}+1)+s_{24}(s_{24}-1)-4s(s+1)]\hbar^2.
\end{align}

		\begin{figure}[t]
   			\centering
            \hskip -0.9cm
   			\includegraphics[width=9.5cm]{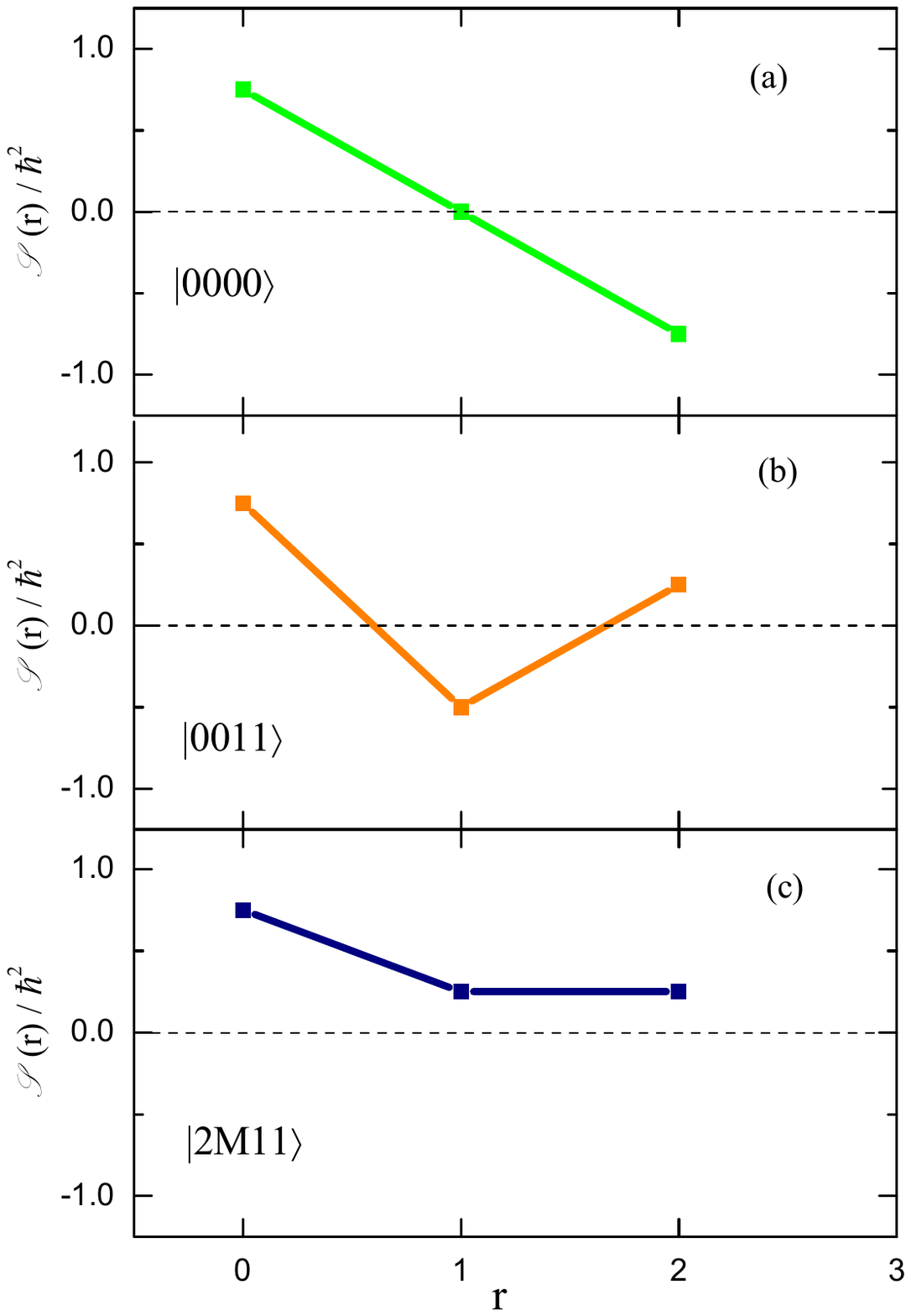} 
            \vskip -0.75cm
     			\caption{Correlation functions for a 4-site ring in the $\ket{S_TMs_{13}s_{24}}$ states as functions of the distance between spins: (a) $\ket{0000}$,  (b) $\ket{0011}$, and (c) $\ket{2M11}$; in the latter case, the quintuplet, they are degenerate in $M=-2,-1,0,1, 2$.
			}
  			\label{fig:corr4}
		\end{figure}

The correlation functions for the different ground states (in different regimes of $\alpha$) are displayed in Fig.\,\ref{fig:corr4}.
Figure \ref{fig:corr4}(a) shows that when the ground state is the singlet given by Eq.\,\eqref{eq:0000otimestxt}, spins on the same sublattice are maximally anticorrelated, since they are in two-spin singlet states; spins on different sublattices, on the other hand, are uncorrelated. 
This should be contrasted with Fig.\,\ref{fig:corr4}(b), for which the overall singlet, Eq.\,\eqref{eq:0011otimestxt}, is made up from two-spin triplets adding to yield $S=0$. 
Indeed, members of the two-spin triplets are positively correlated, but negatively correlated with a spin in the other sublattice; classically the triplets would be antiparallel to each other.
Also noteworthy is the overall decay in the magnitude of $\scrS(r)$.

The correlation function for the quintuplet does not depend on $M$, as a result of this basis satisfying rotational invariance. 
Had we expressed the ground state in terms of $\{\ket{m_1m_2m_3m_4}\}$, the equality $\ave{S_i^zS_{i+r}^z}=(1/4)\ave{S_i^+S_{i+r}^-+S_{i}^-S_{i+r}^+}$ would only hold by averaging each side of this equation over $M$.
Figure \ref{fig:corr4}(c) shows the correlation function for any member of the quintuplet: all spins are positively correlated, in accordance with the classical picture of a ferromagnetic state; another feature distinguishable from the singlet cases is the tendency of $\scrS(r)$ being a constant beyond $r=1$.

\section{Entanglement}
\label{sec:Entangle4}

The possible states for two spins-1/2, $\ket{S_TM}$, may be expressed as a singlet $\ket{00}$ and a triplet $\ket{11}$, $\ket{1-1}$, and $\ket{10}$; while the states with $M\neq 0$ are \emph{separable}, that is $\ket{11}=\ket{\up}\otimes\ket{\up}$ and  $\ket{1-1}=\ket{\dn}\otimes\ket{\dn}$, the $M=0$ states cannot be written as a direct product of single-particle states, since $\ket{10}\!=\!(1/\sqrt{2})[\ket{\!\up}\!\otimes\!\ket{\!\dn}\!+\linebreak\ket{\!\dn}\!\otimes\ket{\!\up}] $ and 
$\ket{00}=(1/\sqrt{2})[\ket{\up}\otimes\ket{\dn}-\ket{\dn}\otimes\ket{\up}]$.
In the latter cases we say the spins are \emph{entangled}: in a bipartite entangled spin state, determination of the state of one part of the system necessarily implies the knowledge of the outcome of measuring the other part; see, e.g.\,Refs.\,\cite{Griffiths18,Gottfried03}.

The above example with the $M=0$ states suggests that the first spin is in a mixed state, since it has probability 1/2 of being either up or down.  
A qualitative measure of the entanglement between spins 1 and 2 is given by the reduced density matrix for one of the spins, obtained by taking the partial trace of the density operator.
For instance, considering $\rho=\ket{10}\bra{10}$, we trace out the second spin,
\begin{equation}
	\tilde{\rho}(1) \equiv \text{Tr}_{2}\ \rho, 
\end{equation}	
 by calculating the matrix elements \cite{Cohen-Tannoudji77},
\begin{equation}
	\langle m_1 | \tilde{\rho}(1) | m_1' \rangle 
		=\sum_{m_2\,=\,\up,\,\dn} \ \langle m_1m_2 \ket{10}\bra{10}  m_1'm_2\rangle,
\end{equation}
with the result $ \tilde{\rho}(1)=\nicefrac{1}{2}\cdot\mathbb{1}$, where $\mathbb{1}$ is the ($2\times 2$ in this case) identity matrix.	
Since $[\tilde{\rho}(1)]^2\neq\tilde{\rho}(1)$, spin 1 is not in a pure state: it is therefore entangled with spin 2. 

One may also use a quantitative measure of entanglement, namely the von Neumann entropy associated with the reduced density operator \cite{Gottfried03},
\begin{equation}
	S(1) = -\sum_{i=1}^2 \lambda_i\ln\lambda_i,
\end{equation}
where $\lambda_1$ and $\lambda_2$ are the two eigenvalues of $\tilde{\rho}(1)$; in the present case, $\lambda_1=\lambda_2=1/2$, so that $S(1)=\ln 2$, which is the maximum entanglement possible for two spins-1/2 \cite{Gottfried03}. 
By contrast, the entropy is zero for separable states such as $\ket{\up\up}$ and $\ket{\dn\dn}$.

In extending these ideas to the present case of 4 spins, we note from the outset that there are three inequivalent ways of partitioning the system, namely (13)-(24), (12)-(34), and (1)-(234); note that the first partition embraces the good quantum numbers $s_{13}$ and $s_{24}$, so one expects features different from the (12)-(34) partition. 
We now consider each of these bipartite cases in turn.

\subsection{Subsystems (13) and (24)}
\label{ssec:13-24}
 
Starting with $\ket{0000}=\ket{00}_{13}\otimes\ket{00}_{24}$, we see that it is obviously separable between two pairs of singlets, (13) and (24). 
This is also manifest by taking the partial trace \cite{Cohen-Tannoudji77} over spins 2 and 4 to obtain the reduced density operator,
\begin{align}
	\tilde{\rho}(13)&=\text{Tr}_{s_{24},m_{24}} \ket{00}_{13}\otimes\ket{00}_{24}\ {_{13}}\bra{00}\otimes {_{24}}\bra{00}\nonumber\\
	&=\ket{00}_{13}\ {_{13}}\bra{00} \ \text{Tr}_{s_{24},m_{24}} \ket{00}_{24}\ {_{24}}\bra{00}\nonumber\\
	&=\ket{00}_{13}\ {_{13}}\bra{00},
\label{eq:Tr2400}	
\end{align}	    
since $\text{Tr}_{s_{24},m_{24}} \ket{00}_{24}\ {_{24}}\bra{00}=1$.
Thus,
\begin{equation}
	[\tilde{\rho}(13)]^2 = \tilde{\rho}(13),
\end{equation}
so that the (13) subsystem is in a pure state, hence separable from (24).
This is hardly surprising, given that the ground state is generated by adding spins 1 and 3 simultaneously with adding 2 and 4.
Accordingly, the eigenvalues of $\tilde{\rho}(13)$ are 1 and 0 (3-fold degenerate) which leads to a vanishing von Neumann entropy.

We now discuss the ground state for $0<\alpha<1/2$, namely $\ket{0011}$, as given by Eq.\,\eqref{eq:0011otimes}.
Unlike  $\ket{0000}$ [Eq.\,\eqref{eq:0000otimes}], one sees by inspection that spins 1 and 3 are entangled with spins 2 and 4. 
Further, following steps similar to those in Eq.\,\eqref{eq:Tr2400} we obtain the reduced density operator,
\begin{equation}
	\tilde{\rho}(13)= \frac{1}{3}\left[\ket{1-1}\bra{1-1}+\ket{10}\bra{10}+\ket{11}\bra{11}\right],
\end{equation}	
which, in the $\ket{s_{13}m_{13}}$ basis, is represented by
\begin{equation}
	\tilde{\rho}(13)= \frac{1}{3}
		\begin{pmatrix} 1 &0 &0&0\\
			0 &1 &0&0\\
			0 &0 &1&0\\
			0 &0 &0&0
		\end{pmatrix},
\end{equation}
and is such that $[\tilde{\rho}(13)]^2\neq\tilde{\rho}(13)$, so that the partitions (13) and (24) are entangled. 
With the eigenvalues of $\tilde{\rho}(13)$ being $\lambda=0$ and $\lambda=1/3$ (3-fold degenerate), the von Neumann entropy becomes 
\begin{equation}
	S(13) = - \sum_{i=1}^4 \lambda_i \ln \lambda_i = \ln 3,
\end{equation}	
yet another signature of entanglement.

In the interval $-1/4 < \alpha <0$, the ground state is the quintuplet $S_T=2$, with 2 separable states, $\ket{2\pm\!211}=\ket{1\pm1}_{13}\otimes\ket{1\pm1}_{24}$, and three entangled states,  
$\ket{2\pm111}$ and $\ket{2011}$; see Eqs.\,\eqref{eq:2111otimes}, \eqref{eq:2011otimes}, and \eqref{eq:2-111otimes}.
The reduced density matrices (now with a subscript $m_{13}$) are 
\begin{equation}
	\tilde{\rho}_{1}(13)= \frac{1}{2}
		\begin{pmatrix} 
			1 &0 &0&0\\
			0 &1 &0&0\\
			0 &0 &0&0\\
			0 &0 &0&0
		\end{pmatrix},
	\tilde{\rho}_{-1}(13)= \frac{1}{2}
		\begin{pmatrix} 
			0 &0 &0&0\\
			0 &0 &0&0\\
			0 &0 &1&0\\
			0 &0 &0&1
		\end{pmatrix},
\end{equation}	
and
\begin{equation}
	\tilde{\rho}_0(13)= \frac{1}{6}
		\begin{pmatrix} 
			1 &0 &0&0\\
			0 &4 &0 &0\\
			0 &0 &1&0\\
			0 &0 &0&0
		\end{pmatrix}.
\end{equation}
The von Neumann entropies for the five states are 
\begin{align}
	S_{\pm2}(13)&=0,\\
	S_{\pm1}(13)&=\ln 2,\\
	S_{0}(13)&=\ln 3 -\frac{1}{3} \ln 2,
\end{align}
so that amongst this quintuplet, the maximally entangled state is the one with $M=0$, corresponding to the largest number of possibilities for $m_{23}$ in the linear combination, Eq.\,\eqref{eq:2011otimes}.

\subsection{Subsystems (12) and (34)}
\label{ssec:12-34}

Let us then investigate what happens when we consider as subsystems the pairs (12) and (34). 
In this case, however, we must express the density operators in the basis of individual spins (see Appendix \ref{app:indiv}) and trace out spins 3 and 4. 
The matrix elements of the reduced density operator, $\tilde{\rho}_{12}$, are
\begin{equation}
	\langle m_1m_2|\tilde{\rho}_{12}|m_1'm_2'\rangle=\sum_{m_3m_4} \langle m_1m_2m_3m_4|\rho|m_1'm_2'm_3m_4\rangle.
\label{eq:ptr34}	
\end{equation}

For $\rho=\ket{0000}\bra{0000}$, we take Eq.\,\eqref{eq:0000indiv} into \eqref{eq:ptr34} to obtain
\begin{equation}
	\tilde{\rho}_{12}=\frac{1}{4}\mathbb{1},
\end{equation}
where $\mathbb{1}$ is the identity matrix ($4\times4$ in this case).	
Since $\tilde{\rho}_{12}^2\neq \tilde{\rho}_{12}$, the subsystem (12) is not pure, meaning that (12) and (34) are entangled.	 
The eigenvalues of $\tilde{\rho}_{12}$ are $\lambda=1/4$ (4-fold degenerate), so that the von Neumann entropy is $S(12)=2\ln 2$.

In the interval $0<\alpha < 1/2$ the density operator is $\rho=\ket{0011}\bra{0011}$ [see Eq.\,\eqref{eq:0001indiv}], and the  trace over $m_3$ and $m_4$ yields 
\begin{equation}
	\tilde{\rho}(12)= \frac{1}{12}
		\begin{pmatrix} 
			1 &0 &0&0\\
			0 &5 &-4&0\\
			0 &-4 &5&0\\
			0 &0 &0&1
		\end{pmatrix},
\end{equation}
whose eigenvalues are $\lambda=1/12$ (3-fold degenerate) and $3/4$, so that the von Neumann entropy is  
\begin{equation}
	S(12) = 2 \ln 2 - \hf \ln 3.
\end{equation}

And, finally, for $-1/4 < \alpha < 0$ the same structure found for $\tilde{\rho}_{m_{13}}$ applies: the $M=\pm2$ states are separable, while the states with $M=0,\pm1$ are entangled. 
Again with the aid of the basis of individual spins, Eq.\,\eqref{eq:ptr34} yields the reduced density operators as	
\begin{equation}
	\tilde{\rho}_{1}(12)= \frac{1}{4}
		\begin{pmatrix} 
			2 &0 &0&0\\
			0 &1 &1&0\\
			0 &1 &1&0\\
			0 &0 &0&0
		\end{pmatrix},
	\tilde{\rho}_{-1}(12)= \frac{1}{4}
		\begin{pmatrix} 
			0 &0 &0&0\\
			0 &1 &1&0\\
			0 &1 &1&0\\
			0 &0 &0&2
		\end{pmatrix},
\end{equation}	
and
\begin{equation}
	\tilde{\rho}_0(12)= \frac{1}{6}
		\begin{pmatrix} 
			1 &0 &0&0\\
			0 &2 &2&0\\
			0 &2 &2&0\\
			0 &0 &0&1
		\end{pmatrix}.
\end{equation}
The associated von Neumann entropies are then
\begin{align}
	S_{\pm2}(12)&=0,\\
	S_{\pm1}(12)&=\ln 2,\\
	S_{0}(12)&=\ln 3 -\frac{1}{3} \ln 2,
\end{align}
which are identical to the ones obtained for the partition (13)-(24). 
This indicates that the entanglement properties of the ground states in this quintuplet are not sensitive to the way the system is partitioned in half.

\subsection{Subsystems (1) and (234)}
\label{ssec:1-234}

Consider now the partition into, say spin 1 and spins 2, 3, and 4. 
The elements of the single-spin reduced matrix, $\tilde{\rho}_{1}$, are then given by
\begin{equation}
	  \langle m_1|\tilde{\rho}_{1}|m_1'\rangle =\sum_{m_2m_3m_4} \langle m_1m_2m_3m_4|\rho|m_1'm_2m_3m_4\rangle.
\label{eq:ptr234}	  
\end{equation}

For the state $\ket{0000}$ we have
\begin{equation}
	  \tilde{\rho}(1)=\frac{1}{4}\mathbb{1},
\end{equation}
where now $\mathbb{1}$ is the $2\times 2$ identity matrix.
Since $[\tilde{\rho(1)}]^2\neq\tilde{\rho}(1)$, spin 1 is entangled with the remaining spins, and the entropy is $S(1)=\ln 2$.

Similarly, for the state $\ket{0011}$ Eq.\,\eqref{eq:ptr234} yields
\begin{equation}
	  \tilde{\rho}(1)=\frac{1}{2}\mathbb{1},
\end{equation}
which leads to the same entropy as for the state $\ket{0000}$.

For the quintuplet of states, we start with $\ket{2\pm211}$:
\begin{equation}
	\tilde{\rho}_{2}(1)= 
		\begin{pmatrix} 
			1 &0 \\
			0 &0
		\end{pmatrix}
		\text{ and }
	\tilde{\rho}_{-2}(1)= 
		\begin{pmatrix} 
			0 &0 \\
			0 &1
		\end{pmatrix},
\end{equation}	
both of which lead to $S_{\pm2}(1)=0$, as it should for a separable state.
For the states $\ket{2\pm111}$, we take Eqs.\,\eqref{eq:indiv2111} and \eqref{eq:indiv2-111} into Eq.\,\eqref{eq:ptr234} to obtain, respectively, 
\begin{equation}
	\tilde{\rho}_{1}(1)= \frac{1}{4}
		\begin{pmatrix} 
			3 &0 \\
			0 &1
		\end{pmatrix}
		\text{ and }
	\tilde{\rho}_{-1}(1)= \frac{1}{4}
		\begin{pmatrix} 
			1 &0 \\
			0 &3
		\end{pmatrix},
\end{equation}	
whose entropies are $S_{\pm1}(1)=2 \ln2 -(3/4)\ln 3$. 
Finally, the reduced density matrix for $\ket{2011}$ becomes $\tilde{\rho}_0(1)=(1/2)\mathbb{1}$, and the associated entropy is $S_0(1)=\ln 2$.

\begin{figure}[t]
   			\centering
            \hskip -1.0cm
   			\includegraphics[width=9.5cm]{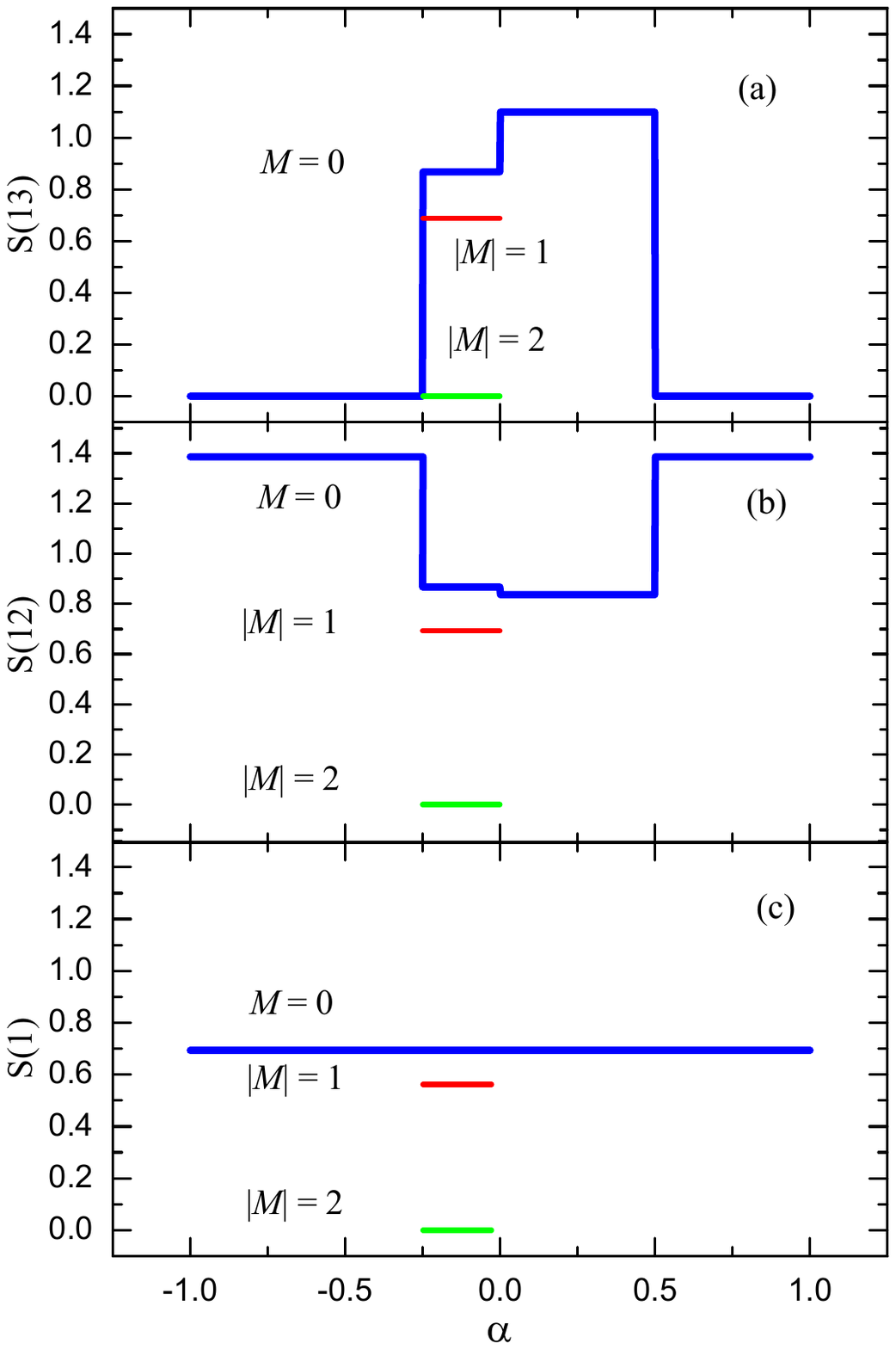} 
            \vskip -1.0cm
     			\caption{Entanglement entropies as functions of $\alpha$: (a) $S(13)$, which measures entanglement between spins on different sublattices, (b) $S(12)$, which measures entanglement between dimers, and (c) $S(1)$, which measures entanglement between a single spin and the remaining ones. 
	In all panels, the entropy depends on $|M|$ in the region $-1/4 < \alpha < 0$.
	}
  			\label{fig:vNS}
		\end{figure}
		

\subsection{Overall discussion of the entropies}
\label{ssec:hierarchy}

Figure \ref{fig:vNS} shows the entropy as a function of $\alpha$ for each partition. 
When comparing the entropies, one should keep in mind that in the FM region, $-1/4<\alpha<0$, there are always three different values of the entropy, such that the entropy vanishes for $|M|=2$ (separable), and increases as $|M|$ decreases.  
When the system is partitioned according to the partial sums, $\Sv_{13}$ and $\Sv_{24}$, one may set up separable states for all $\alpha$, except in the range $0<\alpha<1/2$. 
By contrast, by choosing the (12)-(34) partition, the system is maximally entangled when the ground state is $\ket{0000}$, as given by Eq.\,\eqref{eq:0000otimestxt}: this was to be expected, since spins 1 and 3, as well as 2 and 4, already form singlets themselves.
And, finally, when the partition involves one spin and three spins, we may have at our disposal a uniform entanglement for all values of $\alpha$.   
These findings indicate that coupling q-bits in a controllable way may give rise to a wider range of entanglement outcomes to be explored in quantum computing.

\section{Comments on Larger Systems}
\label{sec:larger}
 
For a 6-site ring a convenient CSCO is $\mathscr{C}_6\equiv\{\Sv_T^2,S^z,\Sv_{14}^2,\Sv_{25}^2,\Sv_{36}^2, \Sv_{2536}^2,\Sv_1^2,\Sv_2^2,\Sv_3^2,\Sv_4^2,\Sv_5^2,\Sv_6^2\}$, where $\Sv_{i,i+3}=\Sv_i+\Sv_{i+3}$, $i=1,2,3$, $\Sv_{2536}\equiv\Sv_{25}+\Sv_{36}$, and $\Sv_T=\Sv_1+\Sv_2+\Sv_3+\Sv_4+\Sv_5+\Sv_6$.
However, it turns out that this CSCO only leads to a direct evaluation of the spectrum for $\alpha=1$, to wit:
\begin{align}
	E_6=\frac{1}{2}&J\hbar^2\left[S_T(S_T+1)- s_{14}(s_{14}+1)-s_{25}(s_{25}+1)\right.\nonumber\\
	&\left.-s_{36}(s_{36}+1) \right],
\label{eq:E6final}		
\end{align}
where $s_{ii+3}=0,1$, $i=1,2,3$.
Note that $E_6$ does not depend on the partial quantum number $s_{2536}$, which, after all, is an arbitrary choice, since one could have used $s_{1425}$ or $s_{1436}$.
The energy is minimized if the state is a total singlet, $S_T=0$, made up by triplet pairs, $s_{14}=s_{25}=s_{36}=1$; we get $E_6=-3J \hbar^2$. 
However, in order to specify the ground state we note that since there is an odd number of triplet pairs, an overall singlet demands that two of them combine into yet another triplet, say $s_{2536}=1$, which, in turn, combines with the remaining triplet, $s_{14}=1$, into an overall singlet. 
While for the 4-spin system the ground state for $\alpha=1$ is made up of singlet pairs, here the ground state is made up of triplet pairs.
Clearly one can still obtain simple solutions for general $\alpha$ by combining states $\ket{m_1m_2\ldots m_6}$ in such way that they are also eigenstates of the translation operator; in this case we end up with a block-diagonal matrix representation of $\calH$ in which the largest block is $4\times 4$.

\begin{figure}[t]
	\vskip 0.5cm
	\centering\includegraphics[scale=0.55]{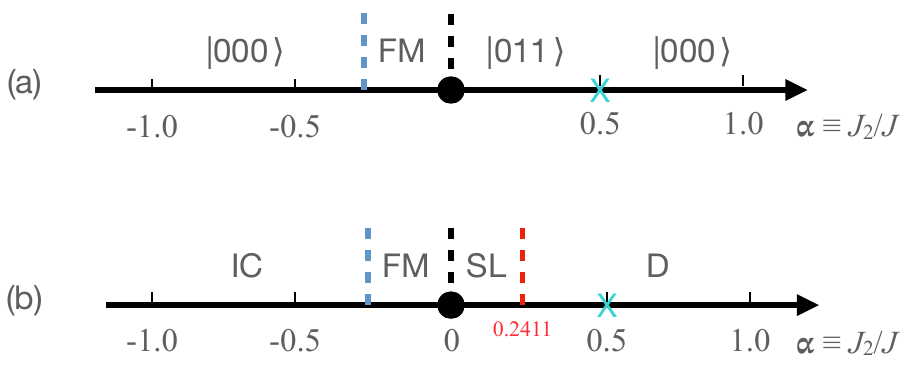} 
	\caption{
 (a) The phase diagram for the 4-site ring, where the kets follow the notation of Fig.\,\ref{fig:Evsa4}, and FM corresponds to the quintuplets $\ket{2M11}$; see text. 
 (b) The phase diagram obtained from different numerical approaches allowing for extrapolations to the thermodynamic limit \cite{Tonegawa89,Okamoto92,Zinke09,Majumdar69}.
 IC stands for incommensurate spiral phase, FM for the saturated ferromagnetic phase, SL for a spin liquid, and D for dimerized; see text.
 }
\label{fig:phase_diag}
\end{figure}

For completeness, we note that numerical studies of this model have been carried out for much larger lattice sizes, $L$, focusing on different quantities \cite{Tonegawa89,Okamoto92,Zinke09} and extrapolating the spectral data for $L\to\infty$; 
for a theory of finite-size scaling involving more than one gap see Ref.\,\ \cite{dosSantos81c}. 
The picture that emerges is that a second order phase transition occurs at $\alpha_c\approx 0.2411$, separating a spin liquid state for $0<\alpha<\alpha_c$, from a dimerized state for $\alpha > \alpha_c$. 
Figure \ref{fig:phase_diag} compares the phase diagrams for the 4-site ring with the one obtained through different numerical methods \cite{Majumdar69,Tonegawa89,Okamoto92,Zinke09}.
Despite the small system size considered here, the two singlet phases we found for $\alpha >0$ may be interpreted as the precursors of the spin liquid and of the dimerized state.
The transitions involving these subtle phases can only be identified on a single system of size $L\gtrsim \xi$, where $\xi$ is the correlation length; however, by examining data for \emph{different sizes}, one may extrapolate to the thermodynamic limit using finite-size scaling theories \cite{Fisher71,Barber83}.
It is also worth stressing that we found that the singlet-from-singlets state and the singlet-from-triplets state cross at $\alpha_\text{MG}=1/2$; this is the so-called Majumdar-Ghosh point \cite{Majumdar69} whose two-fold degeneracy and location do not depend on the system size.
The saturated ferromagnetic phase we found here for $-1/4< \alpha<0$ persists for much larger system sizes, and the first order critical point at $\alpha_F = -1/4$ also does not depend on the system size. 
For $\alpha < -1/4$, however, the $\ket{00}$ ground state we obtained here evolves in the thermodynamic limit to a more subtle incommensurate state, whose classical representation is a spiral spin state. 

Finally, it is worth mentioning that competing interactions along the lines considered here have been invoked to explain the magnetic behavior of materials such as CuGe0$_3$ \cite{Hase93,Castilla95}, SrCuO$_2$ \cite{Matsuda95}, and, more recently, a Szenicsite \cite{Berlie22}.

\section{Conclusions}
\label{sec:concl}

We have considered spins-1/2 placed on each of 4 sites of a ring, coupled through competing nearest-neighbor and next--nearest-neighbor exchange interactions, $J$ and $J_2=\alpha J$, with $J_2>0$ to enforce competition, so that the sign of $\alpha$ is the same as that of $J$. 
This setup allows us to illustrate how addition of more than two angular momenta may simplify the  determination of the spectrum of an interacting system, by sequentially combining pairs of spin angular momenta, forming a total spin quantum number $S_T$.  

We have found that for $\alpha>0$ the ground state is always a singlet, $S_T=0$, but in such way that in the range $0<\alpha<\alpha_\text{MG}=1/2$ this total singlet arises by combining two triplet pairs, while for  $\alpha>\alpha_\text{MG}=1/2$, the ground state is built from two singlet pairs. 
At $\alpha=\alpha_\text{MG}$ the corresponding levels cross each other, so that the ground state is doubly-degenerate.
For $-1/4<\alpha<0$, we found a five-fold degenerate quintuplet of states with $S_T=2$, while for $\alpha<-1/4$ the ground state is a total singlet. 
This procedure to determine the ground states also enormously simplifies the calculation of spin-spin correlation functions, which clearly distinguish the magnetic behaviors. 

This system is also amenable to illustrate several features of the entanglement (von Neumann) entropy to detect changes in the ground state. Again, the simplicity of the solution allows us to establish that the dependence of the entanglement entropy with $\alpha$ is crucially influenced by the way the system is divided into two parts; this may have bearings on the way q-bits are manipulated.

\acknowledgments
The authors are grateful to the Brazilian Agencies, CAPES, CNPq, and FAPERJ for financial support.

\appendix
\section{The split basis}
\label{app:split}

Here we discuss how to express the states $\ket{S_TMs_{13}s_{24}}$ as linear combinations of direct products $\ket{s_{13}m_{13}}\otimes \ket{s_{24}m_{24}}$; we refer to this as \emph{the split basis}.
Since 
\begin{align}
	\langle S_TMs_{13}'s_{24}'|&\left[\ket{s_{13}m_{13}}\otimes\ket{s_{24}m_{24}}\right] =0, \nonumber\\
	& \text{if } s_{13}'\neq s_{13} \text{ or } s_{24}'\neq s_{24},
\end{align}
for a fixed pair $(s_{13},s_{24})$, we may write
\begin{equation}
			\ket{S_TMs_{13}s_{24}}=\sum_{m_{13}+m_{24}=M}
			a_{m_{13},m_{24}}\ket{s_{13}m_{13}}\otimes\ket{s_{24}m_{24}},
\end{equation}
so that the Hilbert space is decomposed into subspaces $\mathcal{E}_{s_{13},s_{24}}$ with dimensions 
$(2s_{13} +1)\cdot(2s_{24} + 1)$; $a_{m_{13},m_{24}}$ are known as the Clebsch-Gordan coefficients \cite{Messiah70v2}, and here will be determined explicitly, for completeness.

We start with the state with the largest $S_T$, namely $S_T=2$, and $M=S_T=2$; see the last column in Table \ref{tab:spins_on_square}.
The only possible correspondence of this state is 
\begin{equation}
	\label{eq:2211otimes}
	\ket{2211}= \ket{11}\otimes\ket{11},
\end{equation}
where from now on we adopt the convention that the first [second] ket on the RHS is relative to the pair (1,3) [(2,4)]. 

Recalling that each one of the ladder operators,  
\begin{equation}
	\label{eq:Spm1324}
		S_T^\pm \equiv S_{13}^\pm+S_{24}^\pm,		
\end{equation}
changes the value of $m$ by one unit, that is  \cite{Messiah70v2} 
\begin{equation}
	S_T^\pm \ket{sm}=\hbar\,\sqrt{s(s+1)-m(m\pm 1)}\,\ket{s\,m\pm1}, 
\label{eq:Spmketsm}	
\end{equation}		
we apply $S_T^-$ to both sides of Eq.\,\eqref{eq:2211otimes} to obtain
		\begin{align}
			S_T^-\ket{2211}&=2\hbar\ket{2111}\nonumber\\
				&=[S_{13}^-+S_{24}^-]\ [\ket{11}\otimes\ket{11}]\nonumber\\
				&=\sqrt{2}\hbar[\ket{10}\otimes\ket{11}+\ket{11}\otimes\ket{10}]
		\end{align}
		from which we extract
		\begin{equation}
			\label{eq:2111otimes}
			\ket{2111}=\frac{1}{\sqrt{2}}[\ket{10}\otimes\ket{11}+\ket{11}\otimes\ket{10}].
		\end{equation}		

Now we further decrease $M$ by one unit, still considering the $S_T=2$ states. 
Following the same steps leading to Eq.\,\eqref{eq:2111otimes} we arrive at
\begin{equation}
	\label{eq:2011otimes}
		\ket{2011}=\frac{1}{\sqrt{6}}
				[\ket{1-\!1}\otimes\ket{11}
				+2 \ket{10}\otimes\ket{10}
				+\ket{11}\otimes\ket{1-\!1}]
\end{equation}	
		
				The states $\ket{2-\!111}$ and $\ket{2-\!211}$ are obtained by flipping all spins in Eqs.\,\eqref{eq:2111otimes} and \eqref{eq:2211otimes}, respectively:		
\begin{equation}
	\label{eq:2-111otimes}
		\ket{2-\!\!111}=\frac{1}{\sqrt{2}}[\ket{10}\otimes\ket{1-\!\!1}+\ket{1-\!\!1}\otimes\ket{10}]	
\end{equation}
and			
\begin{equation}
		\ket{2\!-\!211}=\ket{1-1}\otimes\ket{1-1}.
	\label{eq:2-211otimes}	
\end{equation}		
	
Our next task is to express the states with $S_T=1$ in terms of the split basis. 
Similarly to Eq.\,\eqref{eq:2111otimes}, we may write
\begin{align}
			\label{eq:1111work}
			\ket{1111}&=
			a\ket{10}\otimes\ket{11}+b\ket{11}\otimes\ket{10},
\end{align}	
where the coefficients $a$ and $b$ are chosen by both orthogonalising the state with respect to Eq.\,\eqref{eq:2111otimes},
\begin{equation}
		\braket{2111}{1111}=\frac{1}{\sqrt{2}}(a+b)=0\quad \Rightarrow \quad b=-a,
	\label{eq:scalar2111}
\end{equation}
and normalisation of $\ket{1111}$.
We end up with
\begin{equation}
	\label{eq:1111otimes}
		\ket{1111}=\frac{1}{\sqrt{2}}[\ket{10}\otimes\ket{11}-\ket{11}\otimes\ket{10}],	
\end{equation}
					
		Starting from \eqref{eq:1111otimes}, we repeat the procedure above to generate the states corresponding to $M<1$; we obtain
		\begin{align}
			\label{eq:111Motimes}
			\ket{1011}&=\frac{1}{\sqrt{2}}
				[\ket{1-\!\!1}\otimes\ket{11} -\ket{11}\otimes\ket{1-\!\!1}]\\
			\label{eq:1-111otimes}
			\ket{1-\!\!111}&=\frac{1}{\sqrt{2}}[\ket{10}\otimes\ket{1-\!\!1}-\ket{1-\!\!1}\otimes\ket{10}].	
		\end{align}	
		
We finally arrive at the last state arising from $S_{13}=S_{24}=1$, namely $\ket{0011}$. 
It may be obtained by writing
\begin{equation}
	\ket{0011}=a\ket{11}\otimes\ket{1-1}+b\ket{10}\otimes\ket{10}
	+c\ket{1-1}\otimes\ket{11},		
\end{equation}
with the coefficients being determined by orthogonalisation,
\begin{align}
	&\braket{2011}{0011}=\frac{1}{\sqrt{6}}[c+2b+a]=0\\
	&\braket{1011}{0011}=\frac{1}{\sqrt{6}}[c-a]=0,
\end{align}
together with normalisation; thus,
\begin{equation}
	\label{eq:0011otimes}
	\ket{0011}=\frac{1}{\sqrt{3}}[\ket{1-1}\otimes\ket{11}-\ket{10}\otimes\ket{10}+\ket{11}\otimes\ket{1-1}].
\end{equation}	
		
Now we consider the cases with $S_{13}=1$ and $S_{24}=0$, which leads solely to $S_T=1$. 
We may thus identify
\begin{equation}
	\label{eq:1011otimes}
		\ket{1110}=\ket{11}\otimes\ket{00},	
\end{equation}
to which we successively apply Eq.\,\eqref{eq:Spm1324} to obtain
\begin{align}
	\label{eq:1010otimes}
		\ket{1010}&=\ket{10}\otimes\ket{00}\\
	\label{eq:1-110otimes}
		\ket{1-\!110}&=\ket{1\!-1}\otimes\ket{00}	
\end{align}
The 3 states obtained by  considering $S_{13}=0$ and $S_{24}=1$ are obtained by interchanging the kets on the RHS of Eqs.\eqref{eq:1011otimes}-\eqref{eq:1-110otimes}.

Finally, the total singlet made up by the combination of two partial singlets is
\begin{equation}
	\label{eq:0000otimes}
		\ket{0000}=\ket{00}\otimes\ket{00},	
		\end{equation}
thus completing the 16 states in the split basis.

It is worth noting that some of the states in the split basis are separable, or disentangled (i.e., written in termos of a single direct product), while others are not separable, or entangled.

\section{The individual basis}
\label{app:indiv}

We may also write the eigenstates in terms of 
individual spin states, $\{\ket{m_1m_2m_3m_4}\}$. 
Indeed, since $\ket{11}=\ket{\up\up}$ Eq.\,\eqref{eq:2211otimes} yields
\begin{equation}
	\ket{2211}=\ket{\up\up\up\up};
\label{eq:indiv2211}	
\end{equation}	
Similarly, with $\ket{10}=(1/\sqrt{2})[\ket{\up\dn}+\ket{\dn\up}]$, Eq.\,\eqref{eq:2111otimes} yields
\begin{equation}
	\ket{2111}=\hf[\ket{\dn\up\up\up}+\ket{\up\dn\up\up}+\ket{\up\up\dn\up}+\ket{\up\up\up\dn}],
\label{eq:indiv2111}	
\end{equation}	
and 
\begin{equation}
	\ket{2011}=\frac{1}{\sqrt{6}}[\ket{\!\up\up\dn\dn}+\ket{\!\up\dn\dn\up}+\ket{\!\dn\dn\up\up}+\ket{\!\dn\up\up\dn}+(\ket{\!\up\dn\up\dn}+\ket{\!\dn\up\dn\up})],
\label{eq:indiv2011}	
\end{equation}	
with the remaining states with $S_T=2$ being given by flipping all spins in $\ket{2111}$ and $\ket{2-211}$, that is,
\begin{equation}
	\ket{2-\!111}=\hf[\ket{\up\dn\dn\dn}+\ket{\dn\up\dn\dn}+\ket{\dn\dn\up\dn}+\ket{\dn\dn\dn\up}],
\label{eq:indiv2-111}	
\end{equation}	
and
\begin{equation}
	\ket{2-\!211}=\ket{\dn\dn\dn\dn}.
\label{eq:indiv2-211}	
\end{equation}	

Still with $s_{13}=s_{24}=1$, we have the eigenstates with $S_T=1$, namely
\begin{align}
	\ket{1111} &= \hf[\ket{\up\up\dn\up}+\ket{\dn\up\up\up}-\ket{\up\up\up\dn}-\ket{\up\dn\up\up}]\\
	\ket{1011}&=\frac{1}{\sqrt{2}}[\ket{\dn\up\dn\up}-\ket{\up\dn\up\dn}]\\
	\ket{1-\!111} &= \hf[\ket{\dn\dn\up\dn}+\ket{\up\dn\dn\dn}-\ket{\dn\dn\dn\up}-\ket{\dn\up\dn\dn}],
\end{align}
as well as the one with $S_T=0$,
\begin{equation}
	\ket{0011}\!=\!\frac{1}{\sqrt{3}}[\ket{\up\dn\up\dn}\!+\!\ket{\dn\up\dn\up} 
		-\hf(\ket{\up\up\dn\dn}+\ket{\up\dn\dn\up}+\ket{\dn\up\up\dn}+\ket{\dn\dn\up\up}\!)\!].
\label{eq:0001indiv}		
\end{equation}

For $s_{13}=1$ and $s_{24}=0$, Eqs.\,\eqref{eq:1011otimes}-\eqref{eq:1-110otimes} yield
\begin{align}
	\ket{1110}&=\frac{1}{\sqrt{2}}[\ket{\up\up\up\dn}-\ket{\up\dn\up\up}]\\
	\ket{1010}&=\hf[\ket{\up\up\dn\dn}-\ket{\up\dn\dn\up}+\ket{\dn\up\up\dn}-\ket{\dn\dn\up\up}]\\
	\ket{1-\!110}&=\frac{1}{\sqrt{2}}[\ket{\dn\dn\dn\up}-\ket{\dn\up\dn\dn}],
\end{align}	
while for $s_{13}=0$ and $s_{24}=1$, we have
\begin{align}
	\ket{1101}&=\frac{1}{\sqrt{2}}[\ket{\up\up\dn\up}-\ket{\dn\up\up\up}]\\
	\ket{1001}&=\hf[\ket{\up\up\dn\dn}+\ket{\up\dn\dn\up}-\ket{\dn\up\up\dn}-\ket{\dn\dn\up\up}]\\
	\ket{1-\!101}&=\frac{1}{\sqrt{2}}[\ket{\dn\dn\up\dn}-\ket{\up\dn\dn\dn}].
\end{align}

And, finally, Eq.\,\eqref{eq:0000otimes} leads to
\begin{equation}
	\ket{0000}=\hf[\ket{\dn\dn\up\up}-\ket{\up\dn\dn\up}+\ket{\up\up\dn\dn}-\ket{\dn\up\up\dn}].
\label{eq:0000indiv}	
\end{equation}

\bibliography{rrds-master_bib.bib}

\end{document}